\documentclass[12pt]{article}
\usepackage{epsfig}
\usepackage{axodraw}
\usepackage{epsfig}                            
\usepackage{graphicx}
\usepackage{rotate}
\usepackage{latexsym}
\newcommand\pubnumber{}
\newcommand\pubdate{\today}
\newcommand\hepnumber{hep-ph/0108254}

\def\csumb{Dipartimento di Fisica Teorica, Universit\`a di Torino, Italy\\
INFN, Sezione di Torino, Italy}
\def\support{\footnote{Work supported by the
European Union under contract HPRN-CT-2000-00149.}} 

\def\Title#1{\begin{center} {\Large\bf #1 } \end{center}}
\def\Author#1{\begin{center}{ \sc #1} \end{center}}
\def\Address#1{\begin{center}{ \it #1} \end{center}}

\newcommand\pubblock{\rightline{\begin{tabular}{l} \pubnumber\\
         \pubdate\\ \hepnumber \end{tabular}}}
\newenvironment{Abstract}{\begin{quotation}  }{\end{quotation}}

\makeatletter
\def\section{\@startsection{section}{0}{\z@}{5.5ex plus .5ex minus
 1.5ex}{2.3ex plus .2ex}{\large\bf}}
\def\subsection{\@startsection{subsection}{1}{\z@}{3.5ex plus .5ex minus
 1.5ex}{1.3ex plus .2ex}{\normalsize\bf}}
\def\subsubsection{\@startsection{subsubsection}{2}{\z@}{-3.5ex plus
-1ex minus  -.2ex}{2.3ex plus .2ex}{\normalsize\sl}}

\renewcommand{\@makecaption}[2]{%
   \vskip 10pt
   \setbox\@tempboxa\hbox{\small #1: #2}
   \ifdim \wd\@tempboxa >\hsize     
       \small #1: #2\par          
     \else                        
       \hbox to\hsize{\hfil\box\@tempboxa\hfil}
   \fi}

 \def\citenum#1{{\def\@cite##1##2{##1}\cite{#1}}}
\def\citea#1{\@cite{#1}{}}
 
\newcount\@tempcntc
\def\@citex[#1]#2{\if@filesw\immediate\write\@auxout{\string\citation{#2}}\fi
  \@tempcnta\z@\@tempcntb\m@ne\def\@citea{}\@cite{\@for\@citeb:=#2\do
    {\@ifundefined
       {b@\@citeb}{\@citeo\@tempcntb\m@ne\@citea\def\@citea{,}{\bf ?}\@warning
       {Citation `\@citeb' on page \thepage \space undefined}}%
    {\setbox\z@\hbox{\global\@tempcntc0\csname b@\@citeb\endcsname\relax}%
     \ifnum\@tempcntc=\z@ \@citeo\@tempcntb\m@ne
       \@citea\def\@citea{,}\hbox{\csname b@\@citeb\endcsname}%
     \else
      \advance\@tempcntb\@ne
      \ifnum\@tempcntb=\@tempcntc
      \else\advance\@tempcntb\m@ne\@citeo
      \@tempcnta\@tempcntc\@tempcntb\@tempcntc\fi\fi}}\@citeo}{#1}}
\def\@citeo{\ifnum\@tempcnta>\@tempcntb\else\@citea\def\@citea{,}%
  \ifnum\@tempcnta=\@tempcntb\the\@tempcnta\else
  {\advance\@tempcnta\@ne\ifnum\@tempcnta=\@tempcntb \else\def\@citea{--}\fi
    \advance\@tempcnta\m@ne\the\@tempcnta\@citea\the\@tempcntb}\fi\fi}
\makeatother

\newcommand{\nl}{\nonumber\\}

\newcommand{\lpar}{\left(}                            
\newcommand{\rpar}{\right)}

\newcommand{\bq}{\begin{equation}}                    
\newcommand{\eq}{\end{equation}}
\newcommand{\bqa}{\arraycolsep 0.14em\begin{eqnarray}}
\newcommand{\eqa}{\end{eqnarray}}
\newcommand{\ba}[1]{\begin{array}{#1}}
\newcommand{\ea}{\end{array}}
\newcommand{\ben}{\begin{enumerate}}
\newcommand{\een}{\end{enumerate}}
\newcommand{\bei}{\begin{itemize}}
\newcommand{\eei}{\end{itemize}}
\newcommand{\eqn}[1]{Eq.(\ref{#1})}

\def\Re{\mathop{\operator@font Re}\nolimits}
\def\Im{\mathop{\operator@font Im}\nolimits}
\newcommand{\ord}[1]{{\cal O}\lpar#1\rpar}

\newcommand{\ib}{i}
\newcommand{\asums}[1]{\sum_{#1}}

\newcommand{\wb}{W}

\newcommand{\hk}{K}

\newcommand{\hki}[1]{\phi^{#1}}
\newcommand{\hb}{H}

\newcommand{\fpsiL}{\psi_{_L}}
\newcommand{\fpsiR}{\psi_{_R}}

\newcommand{\fbpsiL}{{\overline{\psi}}_{_L}}
\newcommand{\fbpsiR}{{\overline{\psi}}_{_R}}

\newcommand{\mh}{M_{_H}}

\newcommand{\me}{m_e}

\newcommand{\mhs}{M^2_{_H}}

\newcommand{\mes}{m^2_e}

\newcommand{\gf}{G_{\ssF}}

\newcommand{\sla}[1]{/\!\!\!#1}

\newcommand{\srt}{\sqrt{2}}

\newcommand{\pd}[1]{\partial_{#1}}
\newcommand{\tgen}[1]{\tau^{#1}}

\newcommand{\upar}[1]{u}

\newcommand{\ssF}{{\scriptscriptstyle{F}}}

\newcommand{\ssH}{{\scriptscriptstyle{H}}}

\newcommand{\ssL}{{\scriptscriptstyle{L}}}

\newcommand{\ssR}{{\scriptscriptstyle{R}}}
\newcommand{\ssS}{{\scriptscriptstyle{S}}}
\newcommand{\ssT}{{\scriptscriptstyle{T}}}

\newcommand{\bqas}{\begin{eqnarray*}}
\newcommand{\eqas}{\end{eqnarray*}}

\def\app#1#2 {{\it Acta. Phys. Pol.} {\bf#1},#2}
\def\cpc#1#2 {{\it Computer Phys. Comm.} {\bf#1},#2}
\def\np#1#2 {{\it Nucl. Phys.} {\bf#1},#2}
\def\pl#1#2 {{\it Phys. Lett.} {\bf#1},#2}
\def\prep#1#2 {{\it Phys. Rep.} {\bf#1},#2}
\def\prev#1#2 {{\it Phys. Rev.} {\bf#1},#2}
\def\prl#1#2 {{\it Phys. Rev. Lett.} {\bf#1},#2}
\def\zp#1#2 {{\it Zeit. Phys.} {\bf#1},#2}
\def\sptp#1#2 {{\it Suppl. Prog. Theor. Phys.} {\bf#1},#2}
\def\mpl#1#2 {{\it Modern Phys. Lett.} {\bf#1},#2}
\def\jetp#1#2 {{\it Sov. Phys. JETP} {\bf#1},#2}
\def\fpj#1#2 {{\it Fortschr. Phys.} {\bf#1},#2}
\def\afp#1#2 {{\it Acta.Phys. Polon.} {\bf#1},#2}
\def\err#1#2 {{\it Erratum} {\bf#1},#2}
\def\ijmp#1#2 {{\it Int. J. Mod. Phys} {\bf#1},#2}
\def\nc#1#2 {{\it Nuovo Cimento} {\bf#1},#2}
\def\ap#1#2 {{\it Ann. Phys.} {\bf#1},#2}
\def\cmp#1#2 {{\it Comm. Math. Phys.} {\bf#1},#2}
\def\el#1#2 {{\it Europhys. Lett.} {\bf#1},#2}
\def\hpa#1#2 {{\it Helv. Phys. Acta} {\bf#1},#2}
\def\yf#1#2 {{\it Yad. Fiz.} {\bf#1},#2}
\def\nim#1#2 {{\it Nucl. Instrum. Meth.} {\bf#1},#2}
\def\spz#1#2 {{\it Sov. Pisma Zhetf} {\bf#1},#2}
\def\jetpl#1#2 {{\it JETP Lett.} {\bf#1},#2}
\def\sjnp#1#2 {{\it Sov. J. Nucl. Phys.} {\bf#1},#2}
\def\ptp#1#2 {{\it Progr. Theor. Phys. (Kyoto)} {\bf#1},#2}
\def\rmp#1#2  {{\it Rev. Mod. Phys.} {\bf#1},#2}
\def\zhetf#1#2 {{\it ZhETF} {\bf#1},#2}
\def\prs#1#2 {{\it Proc. Roy. Soc.} {\bf#1},#2}
\def\phys#1#2 {{\it Physica} {\bf#1},#2}

\newcommand{\rcn}{{\rm cn}}
\newcommand{\rdn}{{\rm dn}}
\newcommand{\rsn}{{\rm sn}}
\newcommand{\ram}{{\rm am}}

\def\bfi{\begin{figure}}
\def\efi{\end{figure}}

\begin{document}
\begin{titlepage}
\pubblock

\vfill
\def\thefootnote{\fnsymbol{footnote}}
\Title{Are Constants Constant?} 
\vfill
\Author{Giampiero Passarino\support}
\Address{\csumb}
\vfill
\begin{Abstract}
The prospect of a time-dependent Higgs vacuum expectation value is
examined within the standard model of electroweak interactions.
It is shown that the classical equation of motion for the Higgs field
admits a solution that is a doubly-periodic function of time. 
The corresponding Dirac equation for the electron field is equivalent
to a second order differential equation with doubly-periodic coefficients.
In the limit of very large primitive period of the Higgs background
this equation can be solved in WKBJ approximation, showing
plane-wave solutions with a time-dependent distortion factor which can be made
arbitrarily small.
\end{Abstract}
\vfill
\begin{center}
\end{center}
\end{titlepage}
\def\thefootnote{\arabic{footnote}}
\setcounter{footnote}{0}

\section{Introduction}

The possibility that the fundamental constants of nature might vary with time
has been an object of speculations for many years~\cite{spec}. 
In a modern language, however, one should say that 
the prospect of a time variation in the vacuum expectation value 
(hereafter vev) of the Higgs field seems more plausible than the time 
variation of the Fermi coupling constant $\gf$ or of the electron mass 
$\me$~\cite{Dixit:1988at}.

Changing the vev of the Higgs field has
many physical effects, four of which have astrophysical consequences:
$\gf$ changes, the electron mass $\me$ changes and the nuclear masses and 
binding energies change. All of these effects alter Big Bang nucleosynthesis.
On the other end the change in $\me$ is the only effect relevant for the cosmic
microwave background spectrum~\cite{Kujat:2000rk}.

How much do we know about possible variations of $\me$? We should remember 
that one of the recurring themes in the physics behind the fundamental 
constants is that their values are rarely determined by a direct measurement.
The example of the electron mass illustrates how the information that leads 
to the values of the constants can be indirect and how different paths 
provide redundant constraints on their values. To obtain the best values,  
all of this information are taken into account simultaneously; 
in the approach of the 1998 adjustment~\cite{adju}, the information is divided 
into three
categories: input data, observational equations, and adjusted constants. 
The observational equations are theoretical expressions that give
values of the quantities in the input data category as functions of the 
adjusted constants. The adjusted constants are a suitably chosen set of 
fundamental constants that are determined by the adjustment. The adjustment's 
role is to find the values that best reproduce the input data by means of the 
theoretical expressions. 

There are recent studies that show, for instance, how a change in $\me$ alters 
the CMB fluctuation spectrum. There it is assumed that the variation in $\me$ 
is sufficiently small during the process of recombination so that, one needs
only consider the difference between $\me$ at recombination and $\me$
today, see~\cite{Hannestad:1999xp} and also~\cite{Kaplinghat:1999ry}. 
Furthermore, it has been pointed out~\cite{Kujat:2000rk} that MAP and 
PLANK experiments might be sensitive to variations as small as 
$|\Delta\me/\me| \sim 10^{-2} - 10^{-3}$.

Although all of these considerations look very promising, we are still
missing an important ingredient in the discussion: can a time variation 
in the electron mass (any mass) be made formally consistent with the Standard
Model of strong and electroweak interactions? In other words, do we have to
assume an ad hoc time dependence in these parameters or do we have some 
explicit time variation which is consistent with the mathematical structure 
of the Standard Model? Furthermore, a time-dependent mass is a ill-defined 
concept since there is no stationary state in a time-dependent external field.

In this paper we show that the classical equation of motion for the Higgs 
field in the Minimal Standard Model admits a time-dependent solution which
can be given in terms of a Jacobian elliptic function, therefore a 
doubly-periodic function of time. This solution is explicitly
constructed in Sect. 2. In Sect. 3 we consider the coupling of fermionic 
fields with the Higgs sector and study the effect of their propagation
in a time-dependent Higgs background. The three-momentum ${\bf p}$ is
conserved since the background depends only on time and we can factorize
the usual term, $\exp(i\,{\bf p}\cdot{\bf x})$. In the limit of very large 
primitive period of the Higgs background the Dirac equation for the electron 
field can be solved in WKBJ approximation. The solution can be cast into the
form of a plane-wave with a time-dependent distortion factor which becomes 
arbitrarily small exactly in the limit of infinite primitive period of the 
Higgs background.

\section{A time dependent solution for the Higgs vev}

The relevant part of the Standard Model Lagrangian that we are interested in 
is~\cite{Bardin:1999ak}
\bq
{\cal L}_{\ssH} = - \pd{\mu}K^{\dagger}\pd{\mu}K - \mu^2\,K^{\dagger}K -
                  \frac{1}{2}\,\lambda\,\lpar K^{\dagger}K\rpar^2,
\eq
where $K$ is a complex iso-doublet,
\bqa
\hk = \frac{1}{\srt}
\lpar
\psi+\ib\hki{a}\tgen{a}
\rpar
\lpar
\ba{c}
1 \\ 0
\ea
\rpar,
\eqa
and where we have neglected gauge couplings. We introduce two new quantities 
as follows:
\bq
\mhs = 4\,\frac{\lambda}{g^2}\,M^2, \qquad \mu^2 = \beta - \frac{1}{2}\,
\mhs,
\eq
where $g$ is the $SU(2)$ coupling constant and, for a constant vev, $\mh$ is 
the bare Higgs boson mass. As usual we perform a shift in the $\psi$-field
\bq
\psi = 2\,\frac{M}{g}\,\Phi + \hb.
\eq
The parameter $\beta$ will be adjusted in perturbation theory so that, 
order-by-order, $<0|\hb|0> = 0$.
The well known solution which is at the basis of the so-called spontaneous
symmetry breaking mechanism is $\Phi^2 = 1$. For a constant, non-zero, value 
of $<0|\psi|0>$ $M$ is the bare $\wb$-boson mass and $\mh$ the bare Higgs 
boson mass.
In the present case we allow $\Phi$ to be function of time and arrive at the 
following equation, where according to the usual procedure $\beta = 0$ at
tree level:
\bq
\frac{d^2}{dt^2}\,\Phi - \frac{1}{2}\mhs\,\Phi\,\lpar 1 - \Phi^2\rpar = 0.
\eq
A solution to this equation is given in terms of Jacobian elliptic
functions~\cite{jacef},
\bqa
\Phi(t) &=& N_k\,\rcn\lpar H_kt,k\rpar,  \nl
N^2_k &=& \frac{2\,k^2}{L^2_k}, \quad L^2_k = 2\,k^2 - 1, \quad
H_k = \frac{\mh}{\sqrt{2}\,L_k},
\eqa
and $k$ is any positive real number. Note that we adopt the definition
\bqa
u &=& \int_0^{\phi}\,dt\,\lpar 1 - k^2\,\sin^2t\rpar^{-1/2},  \qquad
\phi = \ram(u), \nl
\rcn(u) &=& \rcn(u,k) = \cos(\ram(u)), \quad \rsn(u) = \rsn(u,k) = 
\sin(\ram(u)),  \nl
\rdn(u) &=& \rdn(u,k) = \Bigl[ 1 - k^2\,\sin^2(\ram(u))\Bigr]^{1/2}.
\label{jef}
\eqa
Such solutions are well known in the literature for both $(1+1)$-dimensional 
and $(3+1)$-dimensional variants of $\phi^4$ theory~\cite{phit}.
The functions $\rcn(z,k),\rdn(z,k)$ have the following properties:
\bqa
\ba{lll}
{\rm periods:} & 4\,{\bf K}, 2\,{\bf K} + 2\,i{\bf K}' &
2\,{\bf K}, 4\,i{\bf K}' \\
{\rm zeros:} & (2\,m+1)\,{\bf K} + 2\,ni\,{\bf K}' &
(2\,m+1)\,{\bf K} + (2\,n+1)\,i\,{\bf K}'  \\
{\rm poles:} & \beta_{mn} = 2\,m{\bf K} + (2\,n+1)i\,{\bf K}'  &  \\
\ea
\eqa
\noindent
where $n,m$ are integers and ${\bf K}(k)$ is the complete elliptic
integral of the first kind 
\bq
{\bf K}(k) = F(\frac{\pi}{2},k), \qquad {\bf K}'(k) = F(\frac{\pi}{2},k'),
\eq
with $k' = (1-k^2)^{1/2}$. Clearly we are interested in solutions with
large values of $k$ since in that limit
\bq
\rcn\lpar H_k,t,k\rpar = \rdn\lpar \tau,\frac{1}{k}\rpar
\sim 1 - \frac{1}{2\,k^2}\sin^2\tau + \ord{\frac{1}{k^4}}, \quad
\tau = k\,H_k\,t, 
\eq
and the vev of the Higgs field is approximately constant with time periodic
fluctuations suppressed by a factor $k^{-2}$. The Lagrangian becomes
\bq
{\cal L}_{\ssH} = - \frac{1}{2}\,\pd{\mu}\hb\pd{\mu}\hb +
2\,\frac{M^2}{g^2}\,\Bigl[ {\dot\Phi}^2 - \lpar \beta - \frac{1}{2}\,
\mhs\rpar\,\Phi^2 - \frac{1}{4}\,\mhs\,\Phi^4\Bigr] + \ord{\hb^2}.
\label{lag}
\eq
This result, and the vev of the $\hb$-field in higher orders of perturbation
theory require some additional comment. There is a total derivative to be 
considered,
\bq
2\,\frac{M}{g}\,k^2\,H^2_k\frac{d}{d\tau}\,\lpar H\,\frac{d\Phi}{d\tau}\rpar.
\eq
Therefore, to quantize the $\hb$-field, we need to consider a time interval 
$\{-\tau_{\ssL},+\tau_{\ssR}\}$
\bq
\tau_{\ssL} = \lpar 2\,m+3\rpar\,{\bf K},  \qquad
\tau_{\ssR} = \lpar 2\,m+1\rpar\,{\bf K},  
\eq
where $m$ is an integer, $4\,{\bf K}$ is the primitive period of 
${\dot \Phi}$ and ${\dot \Phi}(-\tau_{\ssL}) = {\dot\Phi}(\tau_{\ssR}) = 0$. 
For the quantum fluctuations $\hb$ we impose periodic boundary conditions.
At the end the limit $m \to \infty$ will be taken.

In this paper, however, we are not so much interested in the Higgs sector, 
i.e.\ Higgs mass and Higgs self-couplings, or in the problem of time 
dependence of the cosmological constant but rather we concentrate on the 
effect of a time-depended Higgs vev on the {\em masses} of elementary 
particles. For this reason we will analyze in the following section the 
Higgs-fermion Yukawa couplings.

\section{Higgs-fermion interaction}

The relevant piece of the Lagrangian for arbitrary $u,d$ fermion fields will 
be~\cite{Bardin:1999ak}
\bq
{\cal L}_{\ssH-f} = - \fbpsiL\,\sla{\partial}\fpsiL -
\fbpsiR\,\sla{\partial}\fpsiR + \frac{1}{\sqrt{2}}\,g\,\frac{m_d}{M}\,
\fbpsiL K^c d_{_R} - \frac{1}{\sqrt{2}}\,g\,\frac{m_u}{M}\,
\fbpsiL K u_{_R}  + {\mbox h.c.},
\eq
where $\fpsiL$ is a left-handed doublet and $u(d)_{_R}$ are right-handed
singlets of $SU(2)$ with $K^c$ being the charge-conjugate of $K$.
For $\Phi$ constant $m_u,m_d$ are just the bare up, down masses.
For the $(\nu_e,e)$ doublet we find
\bq
{\cal L}_{m} = - {\overline \nu_e}\,\sla{\partial}\nu_e - 
{\overline e}\,\sla{\partial}e - \frac{1}{2}\,g\,\frac{\me}{M}\,\psi^*\,
{\overline e}e.
\eq
Here $\me$ is an arbitrary free parameter with dimension of mass which cannot
be identified with the bare electron mass since the fermion fields are moving
in a time-dependent background. The neutrino remains decoupled from the Higgs 
field while the Dirac equation for the electron becomes
\bq
\lpar \sla{\partial} + \me\,\Phi\rpar\,e = 0.
\label{dirac}
\eq
If we split the field $e$ into upper/lower two-dimensional components, 
$e_+/e_-$, we obtain the following equations:
\bq
i\,\lambda\,\sigma^a\partial_a\,e_{\lambda} + i\,\lambda\,
\partial_t\,e_{-\lambda} + \me\,\Phi\,e_{-\lambda} = 0,
\eq
where $\sigma^a, a= 1,2,3$ are Pauli matrices and $\lambda = \pm 1$.
To find a solution we introduce
\bq
e_{\lambda} = e^{i\,p^ax_a}\,\chi_{_{\lambda}},
\eq
and derive the corresponding equation for $\chi_{_{\lambda}}$,
\bq
\sigma^ap_a\,\chi_{_{\lambda}} - i\,\partial_t\,\chi_{_{-\lambda}} - 
\lambda\me\,\Phi\,\chi_{_{-\lambda}} = 0.
\eq
The above equation is solved by introducing a new set of variables, scalar and 
vector modes: 
\bq 
\chi_{_{\lambda}} = \frac{1}{\sqrt{2}}\,\lpar S_{\lambda} + i\,V^a_{\lambda}\,
\sigma_a\rpar\,\lpar
\ba{c}
1 \\ 0
\ea
\rpar.
\eq
Next we introduce a set of matrices $\sigma^a(p)$, such that
\bqa
p_{\pm} &=& \frac{1}{\sqrt{2}}\,\lpar p_x \mp i\,p_y\rpar, \quad
p = |{\bf p}|,  \nl
U_{ii} &=& N, \quad i=1,2, \qquad U_{12(21)} = \pm\,\sqrt{2}\,N\,\frac{p_{\pm}}
{p+p_z},  \quad N^{-2} = 2\,\frac{p}{p+p_z}.
\eqa
With this definition it is easy to prove~\cite{Passarino:1987bv} that the
matrices
\bq
\sigma^a(p) = U^{\dagger}\,\sigma^a\,U,
\eq
satisfy the following properties:
\bq
\sigma^3(p) = {\bf\sigma}\cdot{\bf p}, \quad \sigma^{1,2}(p) =
{\bf \sigma}\cdot{\hat e}_{1,2},  
\eq
with ${\bf e}_3 = {\bf p}/p$ and
\bq
{\bf e}_i\cdot{\bf e}_j = \delta_{ij},  \qquad
{\bf e}_i \times {\bf e}_j = \varepsilon_{ijk}\,{\bf e}_k,
\eq
Using these matrices we derive a useful decomposition for the vector modes,
\bqa
{\bf V} &=& V_{\ssL}\,{\bf e}_3 + \asums{i=1,2}\,V^i_{\perp}\,{\bf e}_i,  \nl
V^a_{\lambda}\sigma_a &=& V_{\ssL}(\lambda)\,\sigma^3(p) + 
\asums{i=1,2}\, V^i_{\perp}(\lambda)\,\sigma_i(p).
\eqa
By equating the coefficients of ${\bf 1}$ and ${\bf\sigma}(p)$ we obtain
two separate systems of equations relative to the SL (scalar-longitudinal)
and T (transverse) modes:
\bqa
{}&{}&\frac{d}{dt}\,S(\lambda) + i\,\lambda\me\,\Phi\,S(\lambda) - 
p\,V_{\ssL}(-\lambda) = 0,  \nl
{}&{}&\frac{d}{dt}\,V_{\ssL}(\lambda) + i\,\lambda\me\,\Phi\,V_{\ssL}(\lambda) +
p\,S(-\lambda) = 0,  \nl
{}&{}&\frac{d}{dt}\,V^{1,2}_{\perp}(\lambda) + i\,\lambda\me\,\Phi\,
V^{1,2}_{\perp}(\lambda) \pm p\,V^{2,1}_{\perp}(-\lambda) = 0.
\label{aeqs}
\eqa
Solutions of the Dirac equation are classified as follows:
\bq 
\chi^{\ssS\ssL}_{_{\lambda}} = \frac{1}{\sqrt{2}}\,\Bigl[
S(\lambda) + i\,V_{\ssL}(\lambda)\,{\bf e}_3\cdot{\bf\sigma}\Bigr]\,\lpar
\ba{c}
1 \\ 0
\ea
\rpar,
\quad
\chi^{\ssT}_{_{\lambda}} = \frac{i}{\sqrt{2}}\, 
\asums{i=1,2}\,V^i_{\perp}(\lambda)\,{\bf e}_i\cdot{\bf\sigma}\,\lpar
\ba{c}
1 \\ 0
\ea
\rpar.
\eq
A solution to \eqn{aeqs} is obtained by introducing a function $a_{\lambda}$
such that
\bqa
S(\lambda) &=& a_{\lambda}\,f_{\lambda}, \qquad V_{\ssL}(-\lambda) =
{{a_{\lambda}}\over {p}}\,\frac{df_{\lambda}}{dt},  \nl
{{da_{\lambda}}\over {dt}} &=& - i\,\lambda\me\,\Phi\,a_{\lambda},
\label{osys}
\eqa
and similarly for $V^{1,2}_{\perp}(\lambda)$, i.e.
\bq
V^2_{\perp}(\lambda) = a_{\lambda}\,f_{\lambda}, \qquad V^1_{\perp}(-\lambda) =
{{a_{\lambda}}\over {p}}\,\frac{df_{\lambda}}{dt}.
\label{osysp}
\eq
Here the Higgs background is rewritten as
\bqa
\Phi &=& N_k\,\rcn(H_kt,k) = N_k\,\rdn\lpar\tau,\frac{1}{k}\rpar,  \nl
\tau &=& \frac{k\mh\,t}{\sqrt{2}\,L_k} \sim \frac{1}{2}\,\mh\,t,
\quad \mbox{for} \quad k \to \infty,
\eqa
using a relation that is based on the Jacobi's real transformation
\bqa
\rcn\lpar H_kt,k\rpar &=& \rdn\lpar \tau,\frac{1}{k}\rpar,  \nl
\rdn\lpar H_kt,k\rpar &=& \rcn\lpar \tau,\frac{1}{k}\rpar,  \nl
\rsn\lpar H_kt,k\rpar &=& k^{-1}\,\rsn\lpar \tau,\frac{1}{k}\rpar.
\eqa
Moreover, for $k \to \infty$, the following approximations hold:
\bqa
\rcn\lpar\tau,\frac{1}{k}\rpar &\sim&  \cos\tau + \frac{1}{4\,k^2}\,\lpar \tau
- \frac{1}{2}\,\sin2\tau\rpar\,\sin\tau,  \nl
\rdn\lpar\tau,\frac{1}{k}\rpar &\sim&  1 - \frac{1}{2\,k^2}\,\sin^2\tau,  \nl
\rsn\lpar\tau,\frac{1}{k}\rpar &\sim&  \sin\tau - \frac{1}{4\,k^2}\,\lpar \tau
- \frac{1}{2}\,\sin2\tau\rpar\,\cos\tau.
\eqa
With ${\rm r} = \me/\mh$ we obtain a solution for $a_{\lambda}$,
\bqa
a_{\lambda}(\tau) &=& \exp\,\Bigl\{ - 2\,i\lambda {\rm r}\,\arcsin\Bigl[
\rsn\lpar \tau,\frac{1}{k}\rpar\Bigr]\Bigr\}, \nl
{}&\sim& \exp\Bigl\{ - 2\,i\lambda {\rm r}\tau + \ord{\frac{1}{k}}\Bigr\} =
\exp\Bigl\{ - i\lambda\me\,t + \ord{\frac{1}{k}}\Bigr\}.
\label{aeq}
\eqa
The function $f$, therefore, satisfies the following differential equation:
\bq
\frac{d^2f_{\lambda}}{d\tau^2} - 4\,i\lambda {\rm r}\,\rdn\lpar\tau,
\frac{1}{k}\rpar\,\frac{df_{\lambda}}{d\tau} + q^2\,f_{\lambda} = 0,
\label{sode}
\eq
where we have introduced a new parameter
\bq
q^2 = 2\,\frac{L^2_k}{k^2}\,\frac{p^2}{\mhs}.
\eq
The function $\rdn$ is doubly-periodic with periods $2\,{\bf K}$ and 
$4\,i\,{\bf K}'$. Due to the periodicity we can discuss all properties of the 
elliptic function in the so-called fundamental period parallelogram which for
$\rdn$ is $\tau = 2\,\xi\,{\bf K} + 4\,i\eta\,{\bf K}'$ with $0 \le \xi,\eta < 
1$. An irreducible set of poles is given by 
\bq
\beta = \beta_{00} = i\,{\bf K}', \qquad \beta' = \beta_{01} = 3\,i\,{\bf K}',
\eq
with residues $-i$ and $+i$ respectively. Therefore, the singular points of 
the second order differential equation for $f_{\lambda}$ are $\tau = \beta,
\beta'$ and their congruent points. The corresponding exponents are
$0, 1 \pm 4\,\lambda {\rm r}$ for $\beta(\beta')$. We know that \eqn{sode} 
possesses a fundamental set of solutions but, unfortunately, the exponents are 
not unequal integers and, therefore, we cannot apply the Hermite, Picard, 
Mittag-Leffler, Floquet theorem~\cite{jacef} stating that the solutions are 
doubly-periodic functions of second kind and, in general, expressible as
products of ratios of weierstrassian $\sigma$-functions, see 
also~\cite{Stevenson:1957ee}.

We have not been able to find an exact, explicit, solution to \eqn{sode} but 
an approximated one can be given by using the WKBJ method, based on the 
observation that, for $k \to \infty$,
\bqa
{\dot \Phi} &=& - N_kH_k\,\rdn\lpar H_kt,t\rpar\,\rsn\lpar H_kt,k\rpar \sim
- \frac{1}{2\,k^2}\,\,\mh\,\cos \tau\sin \tau,  \nl
\Phi &=& N_k\,\rcn\lpar H_kt,k\rpar \sim 1 - \frac{1}{2\,k^2}\,\sin^2 \tau.
\eqa
Therefore, in this limit, we obtain
\bqa
f^{\pm}_{\lambda}(\tau) &=& \exp\Bigl\{ i\,\int_0^{\tau}\,du\,
\Theta^{\pm}_{\lambda}(u),\Bigr\},  \nl
\Theta^{\pm}_{\lambda}(u) &=& 2\,\lambda\,{\rm r}\,\rdn\lpar u,
\frac{1}{k}\rpar \pm 
\Bigl[ q^2 + 4\,{\rm r}^2\,\rdn^2\lpar u,\frac{1}{k}\rpar\Bigr]^{1/2}.
\label{wkbs}
\eqa
The integrals appearing in \eqn{wkbs} give
\bqa
\int_0^{\tau}\,du\,\rdn\lpar u,\frac{1}{k}\rpar &=& \arcsin\lpar \rsn\lpar
\tau,\frac{1}{k}\rpar\rpar 
\sim \tau - \frac{1}{4\,k^2}\,\lpar \tau - \frac{1}{2}\,\sin 2\,\tau\rpar
+ \ord{k^{-4}},  \nl
{}&{}&  \nl
\int_0^{\tau}\,du\,\Bigl[ q^2 + 4\,{\rm r}^2\,\rdn^2\Bigr]^{1/2} &=&
\int_0^{\tau}\,du\,\Bigl[4\,{\cal R} - h^2\,\sin^2u + 
\ord{k^{-4}}\Bigr]^{1/2},  \nl
{}&=& 2\,\sqrt{{\cal R}}\,E\lpar \tau,\frac{h}{2\,\sqrt{{\cal R}}}\rpar + 
\ord{k^{-4}}  \nl
{}&=&
2\,\sqrt{{\cal R}}\,\tau - \frac{1}{8}\,\frac{h^2}{\sqrt{{\cal R}}}\,\lpar 
\tau - \frac{1}{2}\,\sin(2\,\tau)\rpar + \ord{h^4},
\nl
\eqa
where we have introduced the following parameters,
\bq
{\cal R} = \frac{1}{4}\,q^2 + {\rm r}^2 = \frac{p^2+\mes}{\mhs} +
\ord{k^{-2}}, \quad h = 2\,\frac{r}{k},
\eq
and where $E$ is the elliptic integral of second kind. In the limit $k \to 
\infty$ we also have
\bq
{\cal E}^2 = p^2 + \mes, \quad
{\cal R} = \frac{{\cal E}^2}{\mhs} - \frac{p^2}{2\,\mhs k^2}, \quad
2\,\sqrt{{\cal R}}\,\tau \sim {\cal E}\,t. 
\label{fen}
\eq
To summarize, in WKBJ approximation, we find the following result for
$\Theta$ of \eqn{wkbs},
\bq
\Theta^{\pm}_{\lambda}(u) = \lambda{\rm r}\,\Bigl[2 - \frac{1}{k^2}\,
\sin^2u\Bigr] \pm 2\,\sqrt{{\cal R}}\,\lpar 1 - \frac{h^2}{2\,{\cal R}}\,
\sin^2u\rpar + \ord{k^{-4}}.
\eq
Note that in the product $a_{\lambda}\,f_{\lambda}$ (see \eqn{aeq})  
$\arcsin(\rsn(\tau,1/k))$ drops out and, by interacting with the 
time-dependent Higgs background, the positive(negative) energy plane 
wave-solutions that would correspond to a free electron of mass $\me$,
\bq
\exp\Bigl\{\pm\,i\,{\cal E}\,t\Bigr\},
\eq
receive a distortion factor which can be made arbitrarily small for large 
values of $k$. From \eqn{osys} we derive
\bqa
S^{\pm}(\lambda) &=& a_{\lambda}\,f^{\pm}_{\lambda} = 
\exp\Bigl\{ i\,\int_0^{\tau}\,du\,\Theta^{\pm}_{\lambda}(u),\Bigr\},  \nl
V^{\pm}_{\ssL}(-\lambda) &=& i\,S^{\pm}(\lambda)\,\Bigl\{ \lambda\,\frac{\me}{p}\,
\lpar 1 - \frac{\sin^2\tau}{2\,k^2}\rpar \pm 
\frac{{\cal E}}{p}\,\Bigl[ 1 - \frac{p^2}{4\,{\cal E}^2k^2} -
\frac{\mes}{2\,{\cal E}^2k^2}\,\sin^2\tau\Bigr]\Bigr\}.
\eqa
A similar results holds for the transverse components
$V^{1,2}_{\perp}(\lambda)$. A more accurate version of the approximated 
solution will now be derived. Starting from \eqn{sode} we write
\bq
f = \exp(\theta)\,F, \qquad \theta = 2\,i\lambda {\rm r}\,\ram(\tau),
\eq
with $\ram(\tau)$ defined in \eqn{jef} and where $F$ is a solution of
\bq
\frac{d^2F_{\lambda}}{d\tau^2} + \Bigl[ q^2 +4\,{\rm r}^2\,\rdn^2\lpar\tau,
\frac{1}{k}\rpar -2\,i\lambda \frac{{\rm r}}{k^2}\,\rsn\lpar\tau,
\frac{1}{k}\rpar\,\rcn\lpar\tau,\frac{1}{k}\rpar\Bigr]\,F_{\lambda} = 0.
\eq
The standard WKBJ solution of the above equation, based on the fact that 
$\Phi$ is a slowly varying function of $\tau$, follows by introducing
\bq
Q^2_{\lambda}(u) = q^2 + 4\,{\rm r}^2 - 4\,\frac{{\rm r}^2}{k^2}\,\sin^2u - 
2\,i\lambda\frac{{\rm r}}{k^2}\,\sin u\cos u + \ord{k^{-4}}.
\label{qdef}
\eq
We easily derive a solution for $F$,
\bq
F^{\pm}_{\lambda}(\tau) \sim \frac{r}{Q_{\lambda}(\tau)}\,
\exp\Bigl\{\pm\,i\,\int_0^{\tau}\,du\,Q_{\lambda}(u)\Bigr\},
\eq
where the integral in the exponent gives
\bq
\int_0^{\tau}\,du\,Q_{\lambda}(u) = 2\,\sqrt{{\cal R}}\tau - \frac{1}{k^2}\,
\frac{{\rm r}^2}{\sqrt{{\cal R}}}\,\lpar \tau - \frac{1}{2}\,\sin(2\,\tau)\rpar
+ \frac{i}{8\,k^2}\,\frac{\lambda {\rm r}}{\sqrt{{\cal R}}}\,
\lpar\cos(2\,\tau) - 1 \rpar + \ord{k^{-4}}.
\eq
Note that for $p^2 = 0$ we have an exact solution of \eqn{aeqs}. Let 
$f_{\lambda}$ be any of the functions $S(\lambda), V_{\ssL}(\lambda)$, or
$V^i_{\perp}(\lambda)$. The corresponding equation becomes
\bq
\frac{df_{\lambda}}{d\tau} + 2\,i\lambda r\,\rdn\lpar \tau,\frac{1}{k}\rpar\,
f_{\lambda} = 0,
\eq
with a solution
\bq
f_{\lambda} = \exp\Bigl\{- 2\,i\lambda {\rm r}\arcsin\Bigl[
\rsn\lpar\tau,\frac{1}{k}\rpar\Bigr]\Bigr\} = \exp\Bigl\{- 2\,i\lambda {\rm r}\,
\ram(\tau)\Bigr\},
\eq
with $\ram(\tau)$ defined in \eqn{jef}. Expanding for large values of $k^2$ 
gives
\bq
am(\tau) = \tau - \frac{1}{4\,k^2}\,\lpar \tau - \sin\tau\cos\tau\rpar +
\ord{k^{-4}}.
\eq
Since $2\,{\rm r}\tau \sim \me\,t$ we again recover the correct limit
of the free Dirac equation. We arrive at the same conclusion by writing
\eqn{dirac} as
\bq
\frac{d^2\,e}{dt^2} + i\,\me\,{\dot \Phi}\,\gamma^4\,e + \mes\Phi^2\,e = 0,
\eq
and expanding the spinor $e(t)$
\bq
e = \asums{i=1,4} e^i(t)\,u_i,
\eq
where the $u$ are eigenfunctions of the spin generator. It follows
\bq
\frac{d^2\,e^i}{d\tau^2} + 4\,{\rm r}^2\,\rdn\lpar\tau,\frac{1}{k}\rpar\,e^i
\pm 2\,i\frac{{\rm r}}{k^2}\,\rcn\lpar\tau,\frac{1}{k}\rpar\,
\rsn\lpar\tau,\frac{1}{k}\rpar\,e = 0,
\eq
where the $\pm$ refers to $i=1,2$ and $i=3,4$ respectively. Using \eqn{qdef}
we easily derive, in WKBJ approximation,
\bq
e^{1,2} = \frac{{\rm r}}{Q_-(\tau)}\,\exp\Bigl\{- i\,\int_0^{\tau}\,
du\,Q_-(u)\Bigr\},  \quad
e^{3,4} = \frac{{\rm r}}{Q_+(\tau)}\,\exp\Bigl\{+ i\,\int_0^{\tau}\,
du\,Q_+(u)\Bigr\}.
\eq
In the above result it is understood that $p = 0$ and $\sqrt{\cal R} = r$.

\section{Conclusions}

In this paper we have tried to answer an important question related
to the possibility that the fundamental constants of nature might vary with 
time: can a time variation in the electron mass (any mass) be made formally 
consistent with the structure of the Standard Model of strong and electroweak 
interactions? For an exhaustive study of how a change in $\me$ alters 
physical effects we refer to~\cite{Kujat:2000rk}.

We have shown that the classical equation of motion for the Higgs 
field in the Minimal Standard Model admits a time-dependent solution which
can be given in terms of the Jacobian elliptic function $\rcn$ which is a 
doubly-periodic function of time. Therefore, we have a consistent picture
of a time-dependent vev of the Higgs field. The background is not an external 
field but rather it is derived directly from the equations of motion.

By choosing the primitive period of the Higgs vev large enough we have
a picture of the vacuum that is not in any evident contradiction with
plain experimental evidence. By this we mean that the present value of the
proton-to-electron mass ratio is $\mu = 1836.1526645(57)$ and any
{\em significant} variation of this parameter over a small time interval
is excluded ($\Delta\mu/\mu < 2 \times 10^{-4}$~\cite{Potekhin:1998mf}).

Next we have considered the coupling of fermions with the Higgs sector of
the Minimal Standard Model and have studied the effect of their propagation
in the time-dependent Higgs background. 

We have shown that the Dirac equation for the electron field is equivalent
to a second order differential equation with doubly-periodic coefficients.
In the limit of very large primitive period of the Higgs background
this equation can be solved in WKBJ approximation.

We summarize our findings by saying that the solutions to the Dirac equation 
in the slowly varying scalar-background field are plane-waves with a 
time-dependent distortion factor which becomes arbitrarily small in the limit 
of infinite primitive period of the Higgs vev. Therefore fermion fluctuations 
are almost like plane-waves with three-momentum ${\bf p}$ and mass $\me$ but,
strictly speaking, we have no time-dependent mass and ${\cal E}$ of \eqn{fen}
is not the energy of the state; we rather register a departure from the 
plane-wave shape through a time-dependent factor which, of course, could be 
fitted with a plane-wave solution with an effective $\me(t)$ parameter.

\end{document}